\documentclass{elsart}

\usepackage{times,graphics,epsf,epsfig,subfigure,rotate}

\journal{Nuclear Instruments and Methods B}
\begin{document}

\begin{frontmatter}

\title{A Study of Cosmic Ray Secondaries Induced by the 
\emph{Mir} Space Station Using \mbox{AMS-01}}

\def\r{\rlap,}

\author[madrid]{M.Aguilar}
\author[madrid]{J.Alcaraz}
\author[cern]{J.Allaby}
\author[perugia]{B.Alpat}
\author[geneva,perugia]{G.Ambrosi}
\author[eth]{H.Anderhub}
\author[calt]{L.Ao}
\author[moscow]{A.Arefiev}
\author[geneva]{P.Azzarello}
\author[perugia]{E.Babucci}
\author[bologna,mit]{L.Baldini}
\author[bologna]{M.Basile}
\author[grenoble]{D.Barancourt}
\author[lip,ist]{F.Barao}
\author[grenoble]{G.Barbier}
\author[lip]{G.Barreira}
\author[perugia]{R.Battiston}
\author[mit]{R.Becker}
\author[mit]{U.Becker}
\author[bologna]{L.Bellagamba}
\author[geneva]{P.B\'en\'e}
\author[madrid]{J.Berdugo}
\author[mit]{P.Berges}
\author[perugia]{B.Bertucci}
\author[eth]{A.Biland}
\author[perugia]{S.Bizzaglia}
\author[perugia]{S.Blasko}
\author[milan]{G.Boella}
\author[milan]{M.Boschini}
\author[geneva]{M.Bourquin}
\author[bologna]{L.Brocco}
\author[bologna]{G.Bruni}
\author[grenoble]{M.Bu\'enerd}
\author[mit]{J.D.Burger}
\author[perugia]{W.J.Burger}
\author[mit]{X.D.Cai}
\author[aachenIII]{C.Camps}
\author[eth]{P.Cannarsa}
\author[mit]{M.Capell}
\author[mit]{G.Carosi}
\author[bologna]{D.Casadei}
\author[madrid]{J.Casaus}
\author[florence,bologna]{G.Castellini}
\author[perugia]{C.Cecchi}
\author[ncu]{Y.H.Chang}
\author[hefei]{H.F.Chen}
\author[ihep]{H.S.Chen}
\author[calt]{Z.G.Chen}
\author[kurch]{N.A.Chernoplekov}
\author[ncu]{T.H.Chiueh}
\author[korea]{K.Cho}
\author[ewha]{M.J.Choi}
\author[ewha]{Y.Y.Choi}
\author[as]{Y.L.Chuang}
\author[bologna]{F.Cindolo}
\author[aachenIII]{V.Commichau}
\author[bologna]{A.Contin}
\author[geneva]{E.Cortina-Gil}
\author[geneva]{M.Cristinziani}
\author[coimbra]{J.P.da\,Cunha}
\author[mit]{T.S.Dai}
\author[madrid]{C.Delgado}
\author[mit]{B.Demirk\"oz}
\author[ist]{J.D.Deus}
\author[perugia]{N.Dinu\thanksref{HEPPG}}
\author[eth]{L.Djambazov}
\author[bologna]{I.D'Antone}
\author[iee]{Z.R.Dong}
\author[geneva]{P.Emonet}
\author[hut]{J.Engelberg}
\author[mit]{F.J.Eppling}
\author[turku]{T.Eronen}
\author[perugia]{G.Esposito}
\author[geneva]{P.Extermann}
\author[lapp]{J.Favier}
\author[perugia]{E.Fiandrini}
\author[mit]{P.H.Fisher}
%  R.Flaminio}lapp\ removed by req JP Vialle 28.02.00
\author[aachenIII]{G.Fluegge}
\author[lapp]{N.Fouque}
\author[moscow,mit]{Yu.Galaktionov}
\author[milan]{M.Gervasi}
\author[bologna]{P.Giusti}
\author[milan]{D.Grandi}
\author[eth]{O.Grimm}
\author[iee]{W.Q.Gu}
\author[aachenIII]{K.Hangarter}
\author[eth]{A.Hasan}
\author[mit]{R.Henning\corauthref{cor}\thanksref{lbnl}}
\corauth[cor]{Corresponding author.}
\ead{rhenning@lbl.gov}
\author[lapp]{V.Hermel}
\author[eth]{H.Hofer}
\author[as]{M.A.Huang}
\author[eth]{W.Hungerford}
\author[perugia]{M.Ionica\thanksref{HEPPG}}
\author[perugia]{R.Ionica\thanksref{HEPPG}}
\author[eth]{M.Jongmanns}
\author[hut]{K.Karlamaa}
\author[aachenI]{W.Karpinski}
\author[eth]{G.Kenney}
% ...

% --------------------------------------------------------------------------

\address[madrid]{ Centro de Investigaciones Energ{\'e}ticas, 
     Medioambientales y Tecnol\'ogicas, CIEMAT, E-28040 Madrid,
     Spain\thanksref{CICT}}
\address[cern]{ European Laboratory for Particle Physics, CERN, 
            CH-1211 Geneva 23, Switzerland}
\address[perugia]{ INFN-Sezione di Perugia and Universit\'a Degli 
     Studi di Perugia, I-06100 Perugia, Italy\thanksref{ISA}}
\address[geneva]{ University of Geneva, CH-1211 Geneva 4, Switzerland}
\address[eth]{ Eidgen\"ossische Technische Hochschule, ETH Z\"urich,
     CH-8093 Z\"urich, Switzerland}
\address[calt]{ Chinese Academy of Launching Vehicle Technology, CALT,
  100076 Beijing, China}
\address[moscow]{ Institute of Theoretical and Experimental Physics, ITEP, 
     Moscow, 117259 Russia}
\address[bologna]{ University of Bologna and INFN-Sezione di Bologna, 
     I-40126 Bologna, Italy\thanksref{ISA}}
\address[mit]{Massachusetts Institute of Technology, Cambridge, MA 02139, USA}
\address[grenoble]{ Institut des Sciences Nucleaires, IN2P3/CNRS,
     F-38026 Grenoble, France}
\address[lip]{ Laboratorio de Instrumentacao e Fisica Experimental de 
            Particulas, LIP, P-1000 Lisboa, Portugal}
\address[ist]{ Instituto Superior T\'ecnico, IST,  P-1096 Lisboa, Portugal}
\address[milan]{ INFN-Sezione di Milano, I-20133 Milan, 
     Italy\thanksref{ISA}}
\address[aachenIII]{
 III. Physikalisches Institut, RWTH, D-52056 Aachen, Germany\thanksref{DLR}}  
\address[florence]{ CNR--IROE,   
     I-50125 Florence, Italy}
\address[ncu]{ National Central University, Chung-Li, Taiwan 32054}
\address[hefei]{ Chinese University of Science and Technology, USTC,
      Hefei, Anhui 230 029, China\thanksref{NSFC}}
\address[ihep]{ Institute of High Energy Physics, IHEP, 
  Chinese Academy of Sciences, 
  100039 Beijing, China\thanksref{NSFC}}
\address[kurch]{ Kurchatov Institute, Moscow, 123182 Russia}
\address[korea]{ CHEP, Kyungpook National University, 
     702-701 Daegu, Korea}
\address[ewha]{ Ewha Womens University, 120-750 Seoul, Korea}
\address[as]{ Academia Sinica, Taipei 11529, Taiwan}
\address[coimbra]{  Laboratorio de Instrumentacao e Fisica Experimental de 
            Particulas, LIP, P-3000 Coimbra, Portugal}
\address[iee]{ Institute of Electrical Engineering, IEE, 
  Chinese Academy of Sciences, 100080 Beijing, China}
\address[hut]{ Helsinki University of Technology,
    FIN-02540 Kylmala, Finland}
\address[turku]{ University of Turku, FIN-20014 Turku, Finland}
\address[lapp]{ Laboratoire d'Annecy-le-Vieux de Physique des Particules, 
     LAPP, F-74941 Annecy-le-Vieux CEDEX, France}
\address[aachenI]{
 I. Physikalisches Institut, RWTH, D-52056 Aachen, Germany\thanksref{DLR}}
% ...

% --------------------------------------------------------------------------

\thanks[lbnl]{Now at Lawrence Berkeley National Laboratory, Berkeley,
CA 94720, USA}
\thanks[HEPPG]{Permanent address:  HEPPG, Univ.~of Bucharest, Romania.}
\thanks[stpeters]{Permanent address: Nuclear Physics Institute, 
St. Petersburg, Russia.}
\thanks[NIKHEF]{ Now at National Institute for High Energy Physics, NIKHEF, 
           NL-1009 DB Amsterdam, The Netherlands.}
\thanks[ETH]{Supported by ETH Z\"urich.}
\thanks[DLR]{Supported by the 
Deutsches Zentrum f\"ur Luft-- und Raumfahrt, DLR.}
\thanks[NSFC]{Supported by the National Natural Science Foundation 
of China.}
\thanks[ISA]{Also supported by the Italian Space Agency.}
\thanks[CICT]{Also supported by the Comisi\'on Interministerial de 
Ciencia y Tecnolog{\'\i}a.}

% --------------------------------------------------------------------------

\begin{abstract}
The Alpha Magnetic Spectrometer (\mbox{AMS-02}) is a high energy
particle physics experiment that will study cosmic rays in the 
$\sim 100\,\mathrm{MeV}$ to $1\,\mathrm{TeV}$ range and will be installed 
on the International Space Station (ISS) for at least 3 years. 
A first version of \mbox{AMS-02},
\mbox{AMS-01}, flew aboard the space shuttle \emph{Discovery} from
June 2 to June 12, 1998, and collected $10^8$ cosmic ray
triggers. Part of the \emph{Mir} space station
was within the \mbox{AMS-01} field of view during the four day
\emph{Mir} docking phase of this flight. We have reconstructed an image
of this part of the \emph{Mir} space station using 
secondary $\pi^-$ and $\mu^-$ emissions from primary 
cosmic rays interacting with \emph{Mir}. This is the first time this
reconstruction was performed in \mbox{AMS-01}, and it is important for 
understanding potential backgrounds during the 3 year \mbox{AMS-02} 
mission.
\end{abstract}

\end{frontmatter}

% ----------------------------------------------------------------------

\section{Introduction}
\label{se:Mir_effect}

\begin{figure}
\centerline{
\epsfig{file=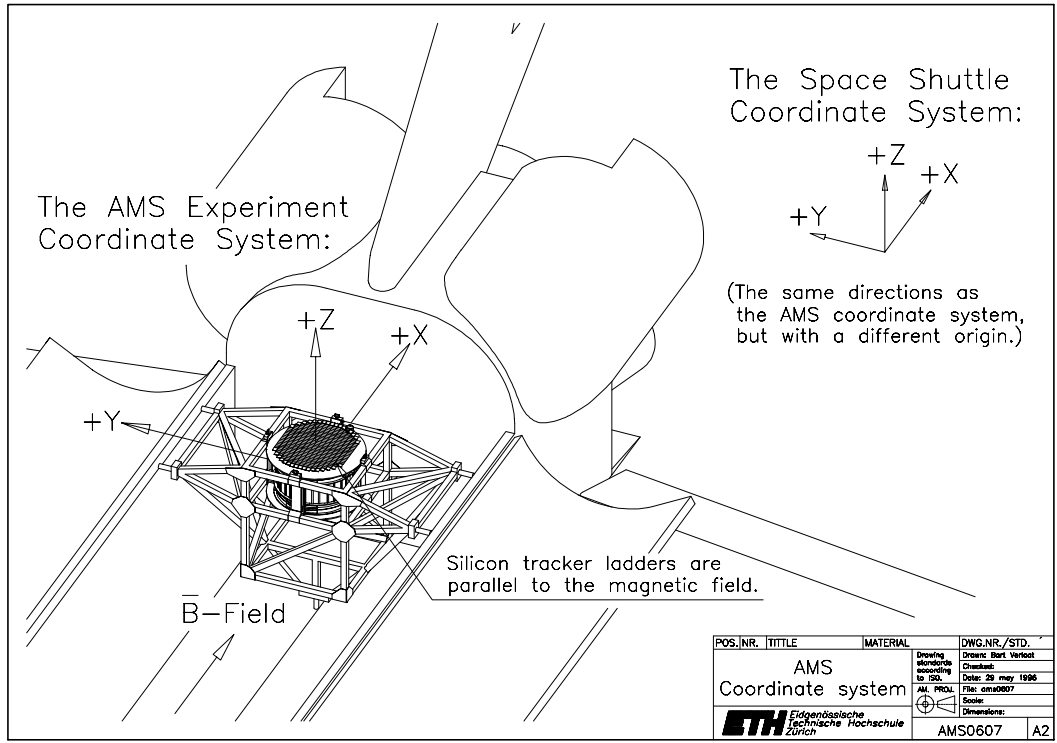, angle=90, width=5.5in}}
\caption{Diagram of \mbox{AMS-01} in the payload bay of \emph{Discovery}.
The space shuttle and \mbox{AMS-01} coordinate systems are shown.}
\label{fig:ams_in_payload}
\end{figure}

\begin{figure}
\centerline{
\epsfysize=6.0in
\epsfbox{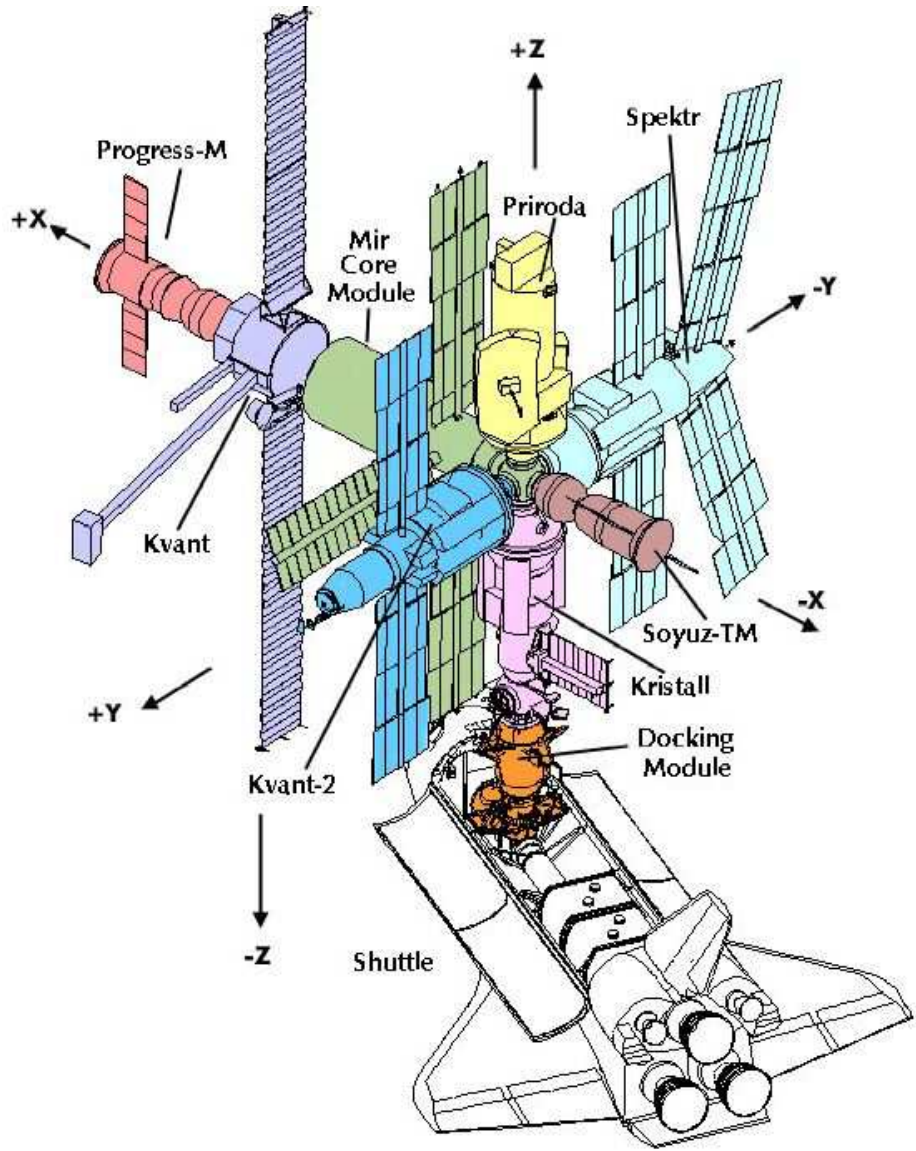}}
\caption{The space station \emph{Mir} and space shuttle shown in a docked
position. The distance from the top of the \emph{Priroda} module to the 
shuttle payload bay is $30 \, \mathrm{m}$. The figure is to scale. 
Note that there is a non-zero angle between the
shuttle $x$-axis (fig.~\ref{fig:ams_in_payload}) and the \emph{Mir} $x$-axis.
Figure courtesy of NASA and taken from~\cite{Mir_Shuttle_WWW}.}
\label{fig:discovery_mir}
\end{figure}

The Alpha Magnetic Spectrometer (\mbox{AMS-02}) is a high energy particle
physics experiment to be installed on the International Space
Station (ISS). It will be mounted on an external truss of the ISS for at 
least three years and collect $\sim 10^{11}$ cosmic ray events in the 
$\sim 100\,\mathrm{MeV}$ to $1\,\mathrm{TeV}$ range. The launch is currently 
scheduled for 2007. 
A first version of the \mbox{AMS-02}
experiment, \mbox{AMS-01}, flew aboard the space shuttle 
\emph{Discovery}  from June 2 to June 12, 1998 
(fig.~\ref{fig:ams_in_payload}). 
Analysis of the 97~million cosmic ray triggers collected by \mbox{AMS-01}
during the flight provided valuable
measurements of cosmic rays in near earth
orbit~\cite{alca99,alca00a,alca00c,alca00b,alca00d}.
\emph{Discovery} was docked with 
the space station \emph{Mir} for 95 of the 235 hours of the 
\mbox{AMS-01} flight (fig.~\ref{fig:discovery_mir}). 
This paper presents a reconstructed image of the part of the \emph{Mir} 
space station that was within the \mbox{AMS-01}
field of view during the flight.
The image is generated from secondary $\pi^-$ and $\mu^-$ emissions
detected by \mbox{AMS-01} that were produced by primary cosmic rays 
interacting with \emph{Mir}. 

% -----------------------------------------------------------------------------

\section{The \mbox{AMS-01} Experiment}
\label{se:AMSexp}

\begin{figure}
\centerline{
\epsfysize=5.5in
\epsfbox{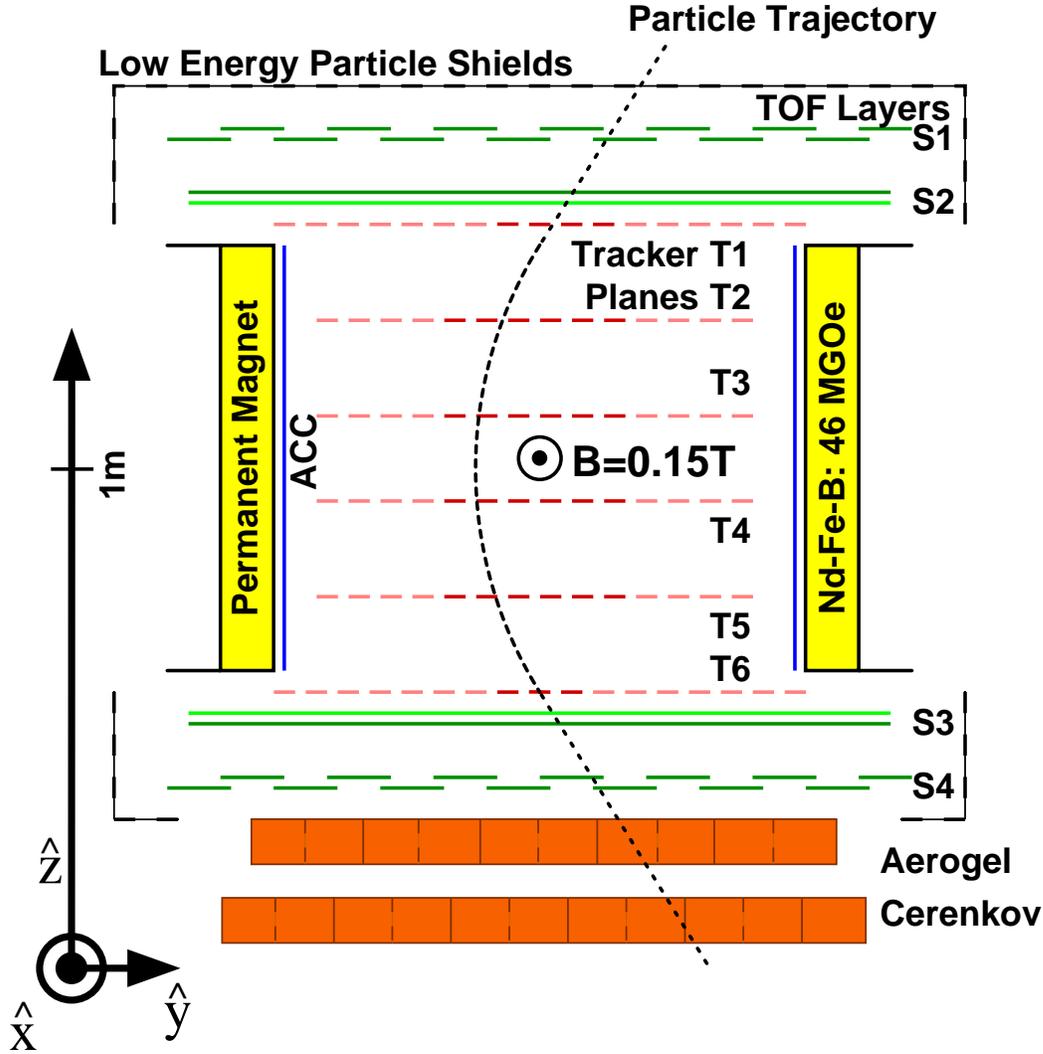} }
\caption{A schematic cross section of the \mbox{AMS-01} detector showing 
the subdetectors.}
\label{fig:AMS_schematic}
\end{figure}

A schematic of the \mbox{AMS-01} detector is shown in 
fig.~\ref{fig:AMS_schematic}. 
\mbox{AMS-01's} core components were a time-of-flight hodoscope (TOF)
that determined the cosmic ray's velocity ($\beta=v/c$), 
and a silicon microstrip tracking system inside 
a permanent magnet that measured particle rigidity ($R$) and charge sign. 
The TOF resolution was $\delta \beta / \beta \approx 3\%$ and 
the tracker rigidity range was $100\,\mathrm{MV}$ to $200\,\mathrm{GV}$
for protons
with an optimal resolution of $\delta R/R \approx 7\%$ at 
$10\,\mathrm{GeV}/c$. Tracker resolution was limited by 
multiple scattering and the bending power of the magnet at low and high
rigidities respectively.
Energy deposits in the TOF paddles and
tracker silicon sensors determined the particle's charge, and 
the probability for charge misidentification between $Z=1$ and $Z=2$ 
particles was estimated to be $<10^{-4}$. An 
anti-coincidence counter (ACC) inside the barrel of the magnet vetoed 
events with secondary particles that missed the TOF detector. 
An Aerogel Threshold \u{C}erenkov detector provided 
additional particle identifying capability. The trigger for recording
an event consisted of a coincidence between the upper and
lower TOF planes and an anticoincidence with the ACC scintillators. 
Reference~\cite{agui02} has a detailed description of the
\mbox{AMS-01} hardware and triggering scheme.

Ionization charge from energy deposits of cosmic rays traversing the 
TOF scintillators and silicon track sensors
were digitized and recorded as hits by the readout electronics.
An offline analysis program reconstructed events from these hits after 
the flight. It performed a three dimensional linear $\chi^2$ fit to 
the time measurements from the TOF and the reconstructed pathlength of the
cosmic ray trajectory in the detector. 
The inverse slope of the fit yielded the cosmic ray's velocity ($\beta$). 
The cosmic ray's trajectory inside the tracking volume (magnet) was 
reconstructed from the hits in the tracker by a
sophisticated tracking algorithm described in~\cite{hart84}. 
It included the effects of multiple 
scattering and the inhomogeneous magnetic field. The tracking algorithm
yielded 5 quantities, $(x,y,\theta,\phi,1/R)$, where ($x,y$) is the 
impact point on the first tracker plane the particle encounters, 
$\theta$ and $\phi$ describes the incident direction of the particle 
relative to the \mbox{AMS-01} coordinate system, and $R$ is the rigidity. 
The particle mass was determined from $R$ and $\beta$ via:
\begin{equation}
\label{eq:mpbrelation}
m = \frac{|Z|R}{c} \sqrt{\beta^{-2}-1},
\end{equation}
where $p=R$ for $Z=1$ particles.
  
% -----------------------------------------------------------------

\section{\emph{Mir} as ``Seen'' by \mbox{AMS-01}} 
\label{se:results}

\begin{figure}
\centerline{
\epsfysize=4.0in
\epsfbox{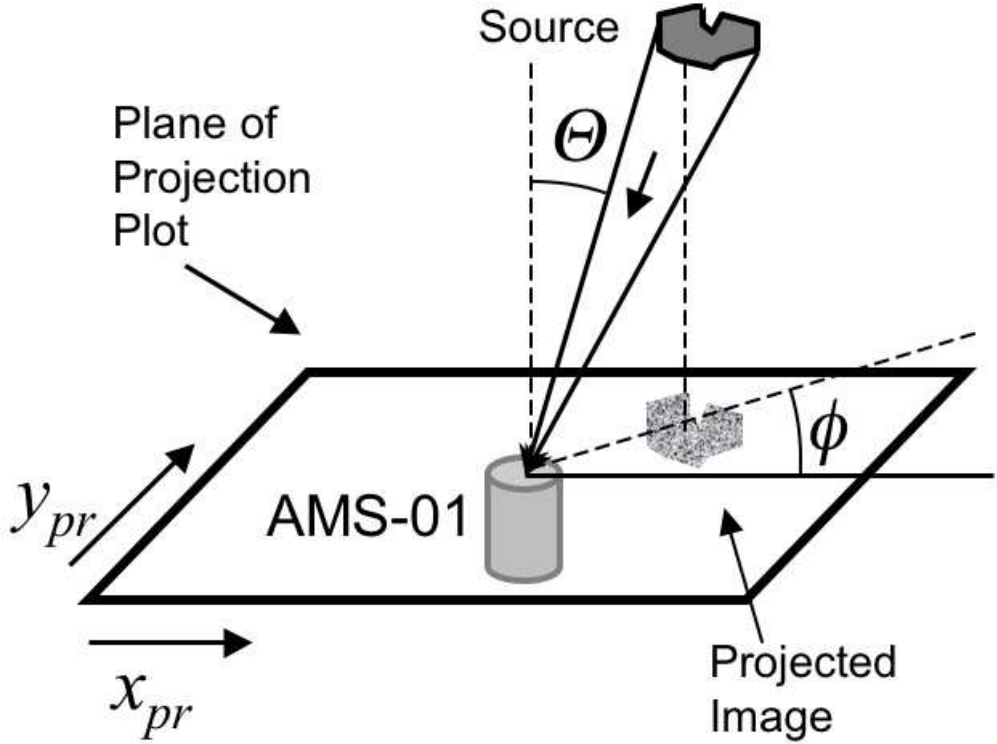}}
\caption{A schematic illustrating how the arrival directions
of incoming cosmic rays are projected onto a plane such that
images of co-moving cosmic ray sources can be created.}
\label{fig:proj_explainer}
\end{figure} 
  
The precision silicon tracker
determined the incident direction of cosmic rays to better than a degree.
The incident directions of cosmic rays were binned 
according to a projection on an $x-y$ plane, such that an 
``image'' is generated (fig.~\mbox{\ref{fig:proj_explainer}}).
Regions of the sky overhead to \mbox{AMS-01} were projected 
one-to-one to
an $x-y$ plane using the standard transformation of arrival direction:
\begin{eqnarray*}
x_{pr} & = & -\sin \theta \: \cos \phi \\
y_{pr} & = & \sin \theta \: \sin \phi,
\end{eqnarray*}
where $\theta$ is the polar angle, and $\phi$ is the azimuthal angle of 
the incoming cosmic ray. 

\begin{figure}
\centerline{
\begin{tabular}{c}
\subfigure{
\epsfysize=3.6in
\epsfbox{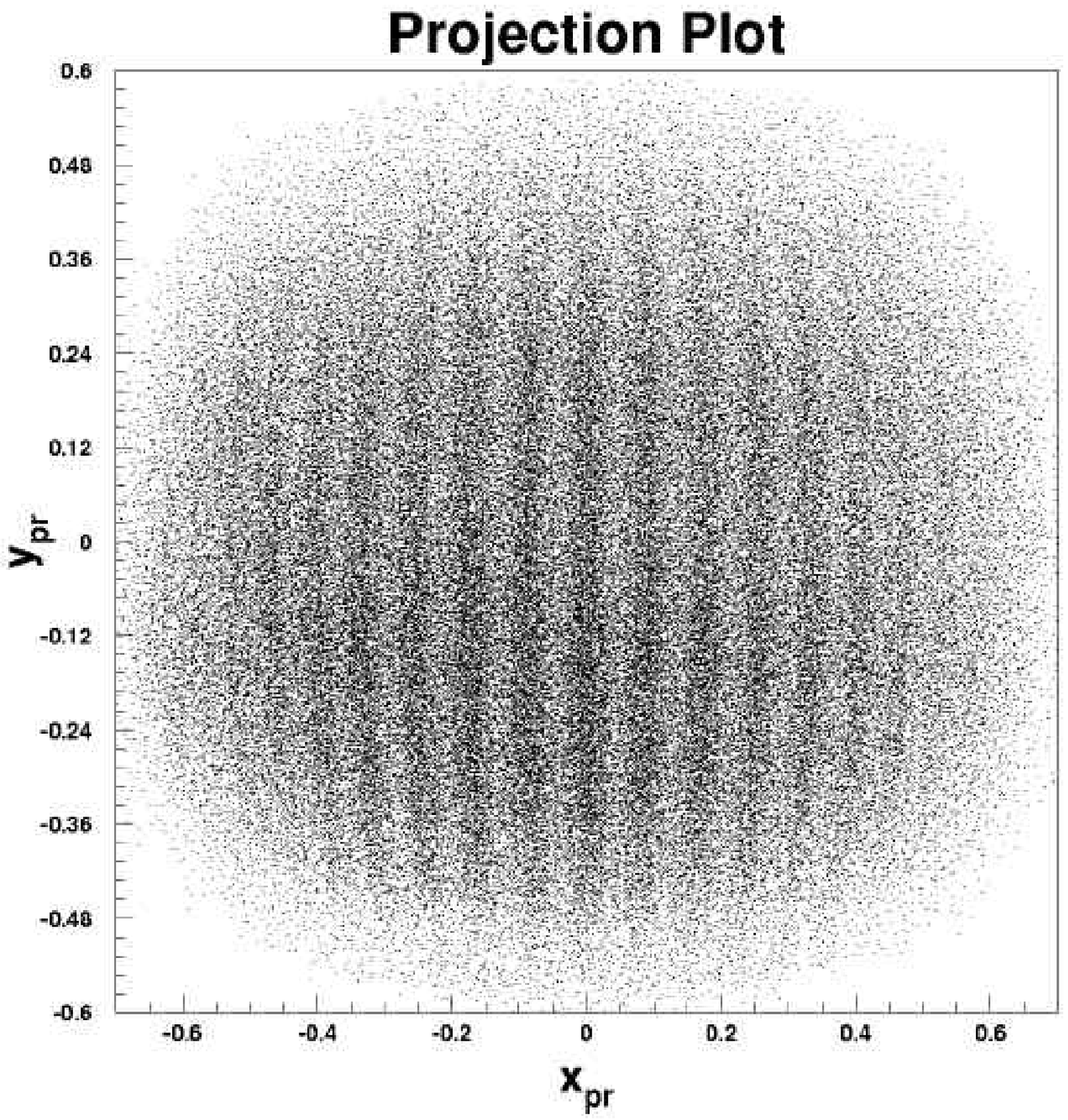}}\\
\subfigure{
\epsfysize=3.6in
\epsfbox{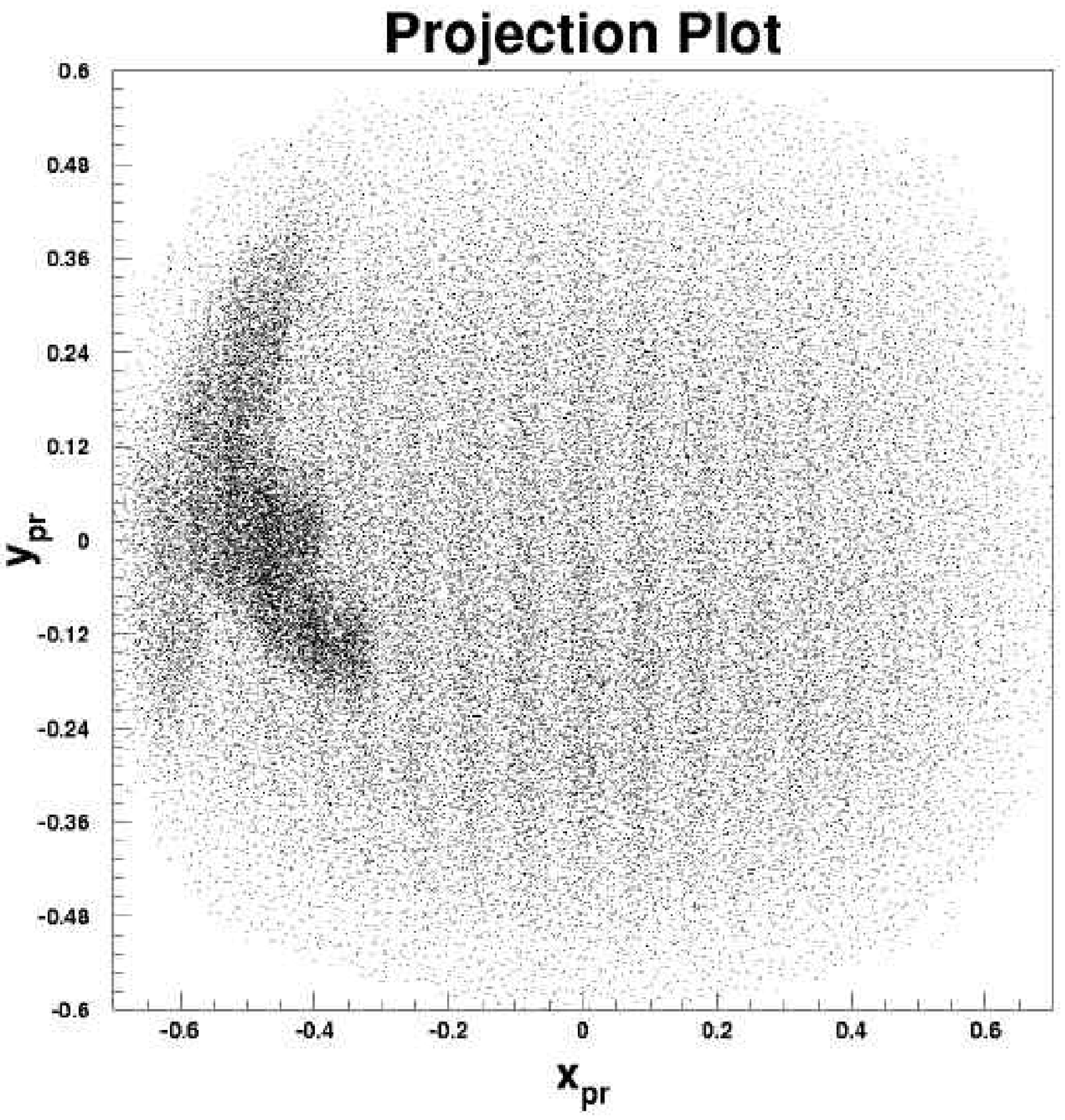}}
\end{tabular}}
\caption{Projection plots for downward-going  particles with 
measured charge $Z=-1$ when \emph{Discovery} was not docked (top) and 
docked (bottom) with \emph{Mir}. Note the relative 
excess of events from a clearly defined region during the
\emph{Mir} docking phase on the left side of the projection plot.
The number of events in the top and bottom plots are $674 000$ and 
$647 000$ respectively. The corresponding rates are $1.42\,\mathrm{s}^{-1}$
(top) and $1.87\,\mathrm{s}^{-1}$ (bottom). 
The rate of events from the excess flux
region on the bottom plot is $0.064\,\mathrm{s}^{-1}$ during the docking phase
and $0.010\,\mathrm{s}^{-1}$ outside the docking phase. The excess flux region
contains $22000$ $Z=-1$ events. 
The vertical stripes are explained in the text.}
\label{fig:zn_nc_down}
\end{figure} 

Fig.~\ref{fig:zn_nc_down} compares projection plots of $x_{pr}$ vs. 
$y_{pr}$ for downward-going $Z=-1$ events when \emph{Discovery} 
was docked and not docked with \emph{Mir}. During the docking phase
the precession of \emph{Mir} and \emph{Discovery} caused \mbox{AMS-01}
to cover most of the sky. Hence, in the absence of a co-moving source, we
would expect a uniform distribution of incident directions for cosmic 
rays, convoluted with the decreasing acceptance of \mbox{AMS-01} at larger
polar angle. However, we observe a significant excess of events from a 
specific
region on the left of the \emph{Mir} docking phase projection plot. This 
region covered $0.7\,\%$ of the \mbox{AMS-01} acceptance. 

As we will show, this excess is due to the \emph{Mir} space station.
The vertical stripes are events that did not have 
reconstructible tracks in the non-bending plane of the tracker.  
The track in the non-bending plane was recovered using the spatial data from
the TOF hodoscope. Hence, the vertical stripes are artifacts of the 
reconstruction and not physical.

\begin{figure}
\centerline{
\begin{tabular}{c}
\subfigure{
\epsfysize=3.6in
\epsfbox{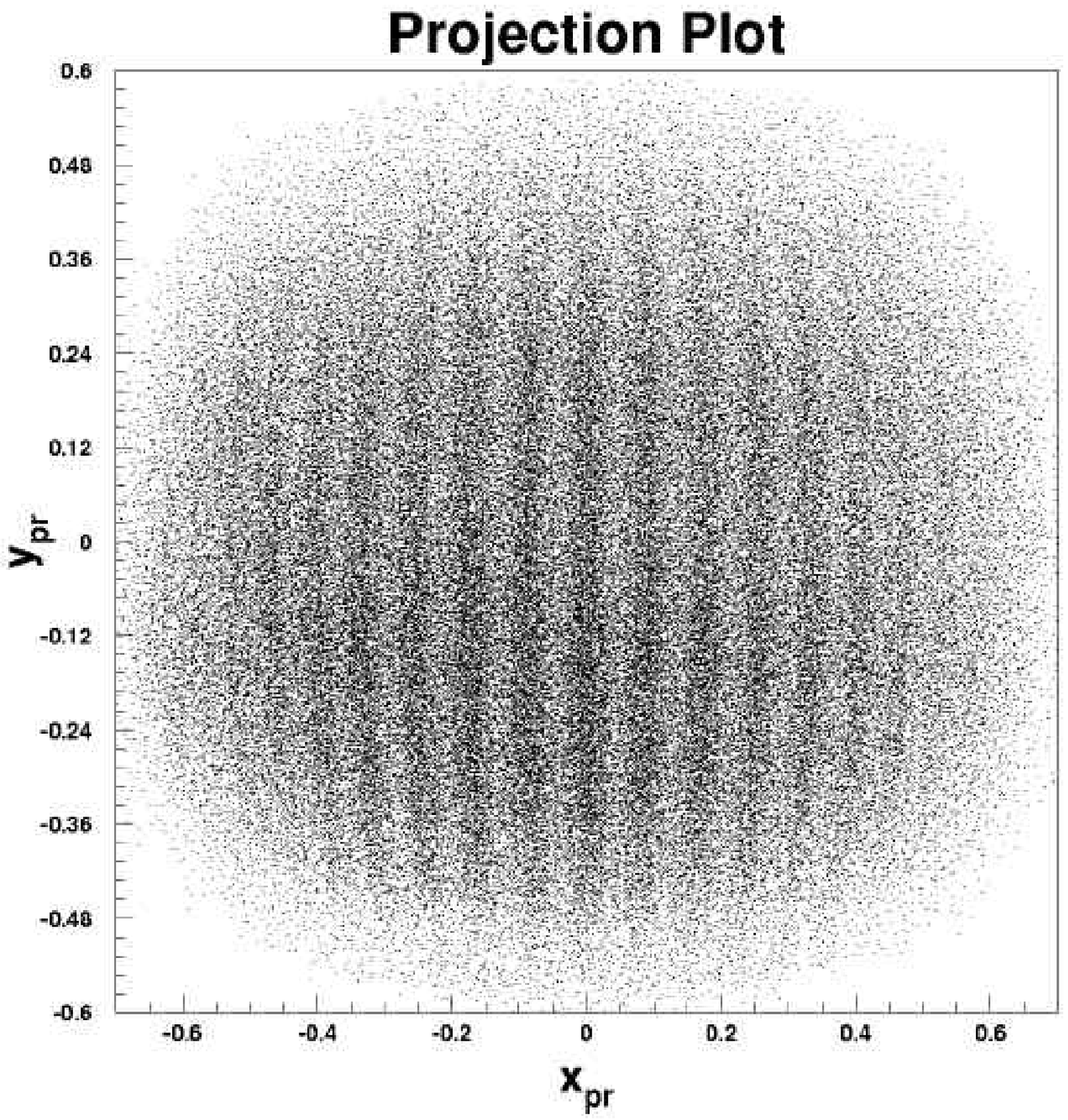}}\\
\subfigure{
\epsfysize=3.6in
\epsfbox{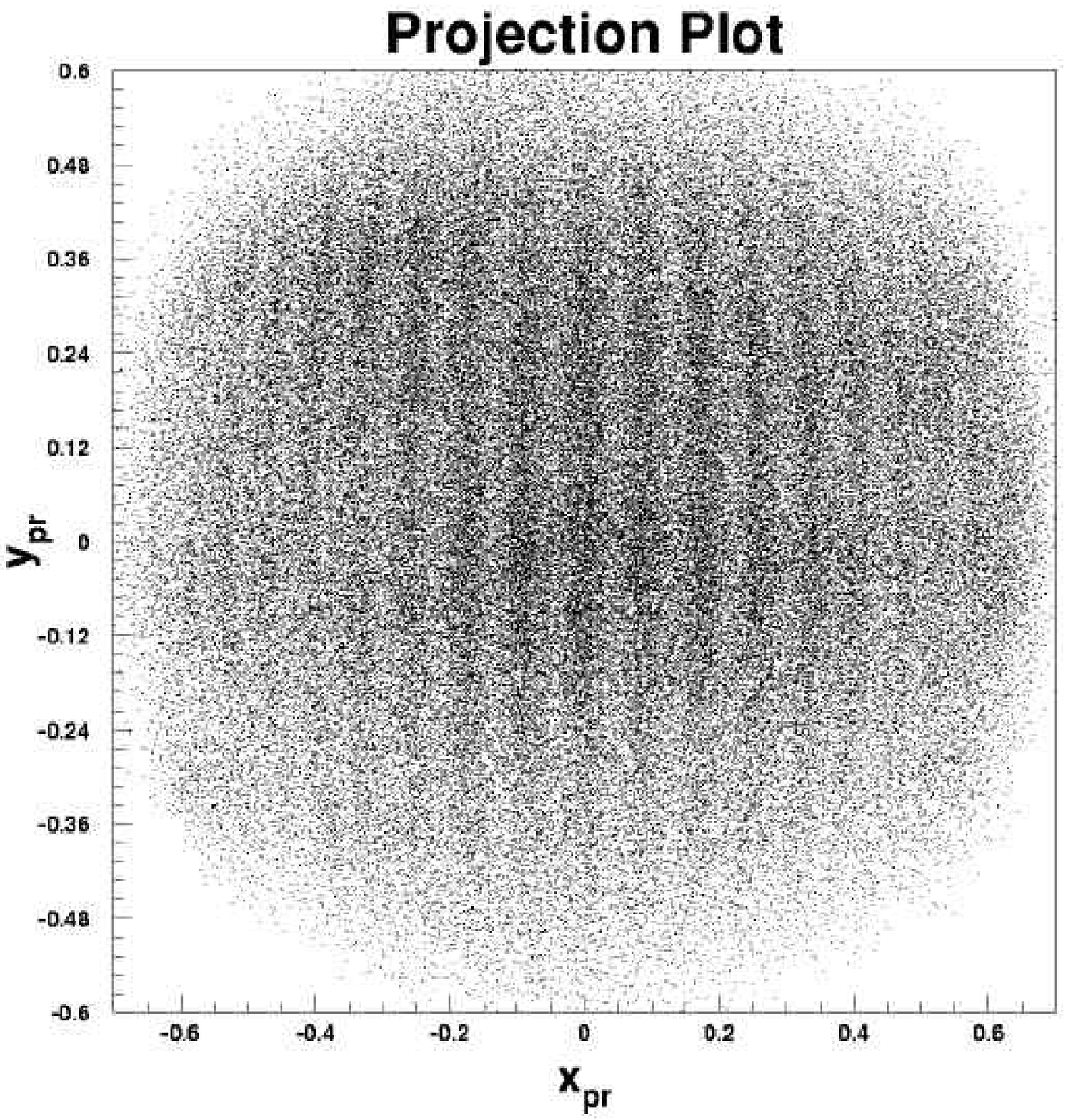}}
\end{tabular}}
\caption{Projection plots for upward-going particles with 
measured $Z=-1$
when \emph{Discovery} was not docked (top) and docked (bottom) with 
\emph{Mir}. No excess event regions are present during the \emph{Mir} 
docking phase.
The plot on the top has \mbox{$538\;000$} events and the one on 
the bottom has \mbox{$389\;000$}. The corresponding event rate is 
$1.1\,\mathrm{event.s}^{-1}$.}
\label{fig:zn_nc_up}
\end{figure} 

Fig.~\ref{fig:zn_nc_up} shows projection plots for upward-going 
particles. No excess is observed in this upward-going sample, although 
faint structures do appear, possibly due to bulkheads and support members 
in the space shuttle airframe and the lower part of the Unique Support 
Structure of \mbox{AMS-01}. 

\begin{figure}
\centerline{
\epsfysize=4.5in
\epsfxsize=5.0in
\epsfbox{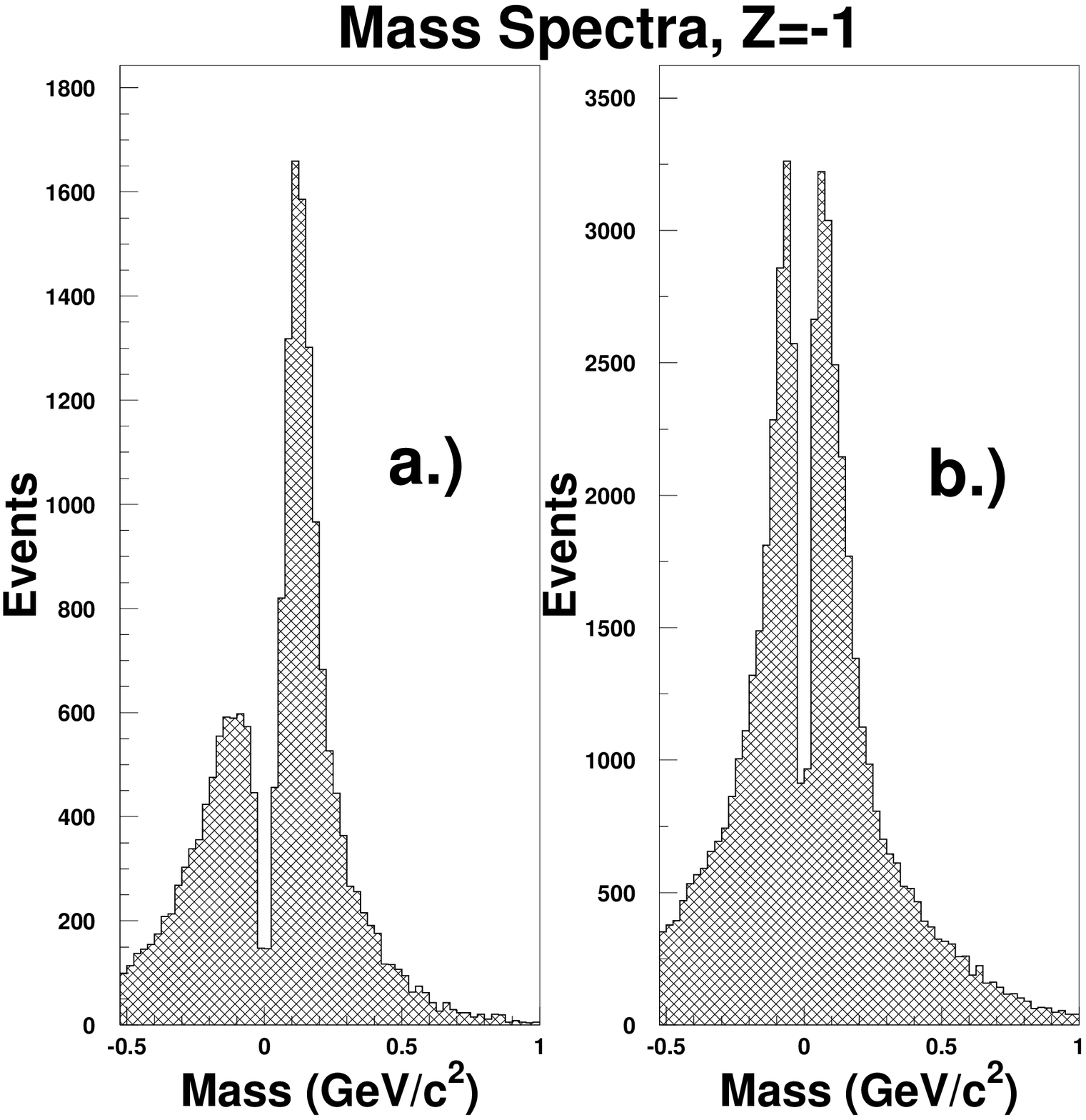}}
\caption{Mass spectra for downward-going $Z=-1$ events from the excess 
flux region (a) and not from the excess flux region (b)
in fig.~\ref{fig:zn_nc_down}. Note the large excess 
of events with masses consistent with $\mu^-$ and $\pi^-$ from the excess 
region. The meaning of negative masses are explained in the text.}
\label{fig:mir_zn_mass}
\end{figure} 

Cuts were applied to improve the mass determination of impinging cosmic rays.
They required agreement between the particle trajectories computed from 
the TOF and tracker, imposed
upper limits on the $\chi^2$ of the track and velocity fits, and rejected
events with bad tracker strips, TOF paddles, and spurious hits.
These cuts rejected events that suffered hard scattering or interactions in
the detector material.  
The resulting mass spectra for $Z=-1$ events are shown in 
fig.~\ref{fig:mir_zn_mass}.  
Events that have measured $\beta > 1$ due to the finite resolution of
the TOF had a ``mass'' computed, as per eq.~\ref{eq:mpbrelation}, with a 
transformed value of velocity: $1/\beta' = 2-1/\beta$. 
These events were tagged by assigning their computed mass a negative sign.
Electrons have $\beta=1$ within the TOF resolution at the energies of 
interest here, hence we expect the measured mass for $e^-$ to be  
distributed symmetrically around $m=0\,\mathrm{GeV}/c^2$.
This distribution is indeed observed in the mass spectra from cosmic rays 
not originating from
the excess on the projection plot (fig.~\ref{fig:mir_zn_mass}.b). This is 
not surprising, since $Z=-1$ cosmic rays are dominated by $e^-$. 
However, there is a large excess of 
events in the \mbox{$0.1-0.2\,\mathrm{GeV}/c^2$} range from the flux excess 
region. A probable source of this excess is $\pi^-$ and $\mu^-$ produced by 
high energy cosmic ray nuclei, primarily protons and Helium, interacting 
hadronically with atomic nuclei in an object in the vicinity of the 
shuttle. The cosmic ray proton flux is $\sim 100$ times that of 
the $Z=-1$ electron flux, hence spallation products from cosmic ray 
protons could contribute significantly to the $Z=-1$ flux.

The mass-resolved $\mu^-$ and $\pi^-$ events shown in the mass histograms
have momentum in the $0.2-0.4\,\mathrm{GeV/}c$ range.
This indicates that the co-moving 
source of the excess cannot be further than several hundred
meters, since for $\mu^-$ and $\pi^-$ we have $c\tau = 660\,\mathrm{m}$ 
and $c\tau=7.8\,\mathrm{m}$ respectively. 

% -----------------------------------------------------------------

\section{Discussion}
\label{se:mir_confirm}

\begin{figure}
\centerline{
\subfigure{
\epsfysize=5in
\epsfbox{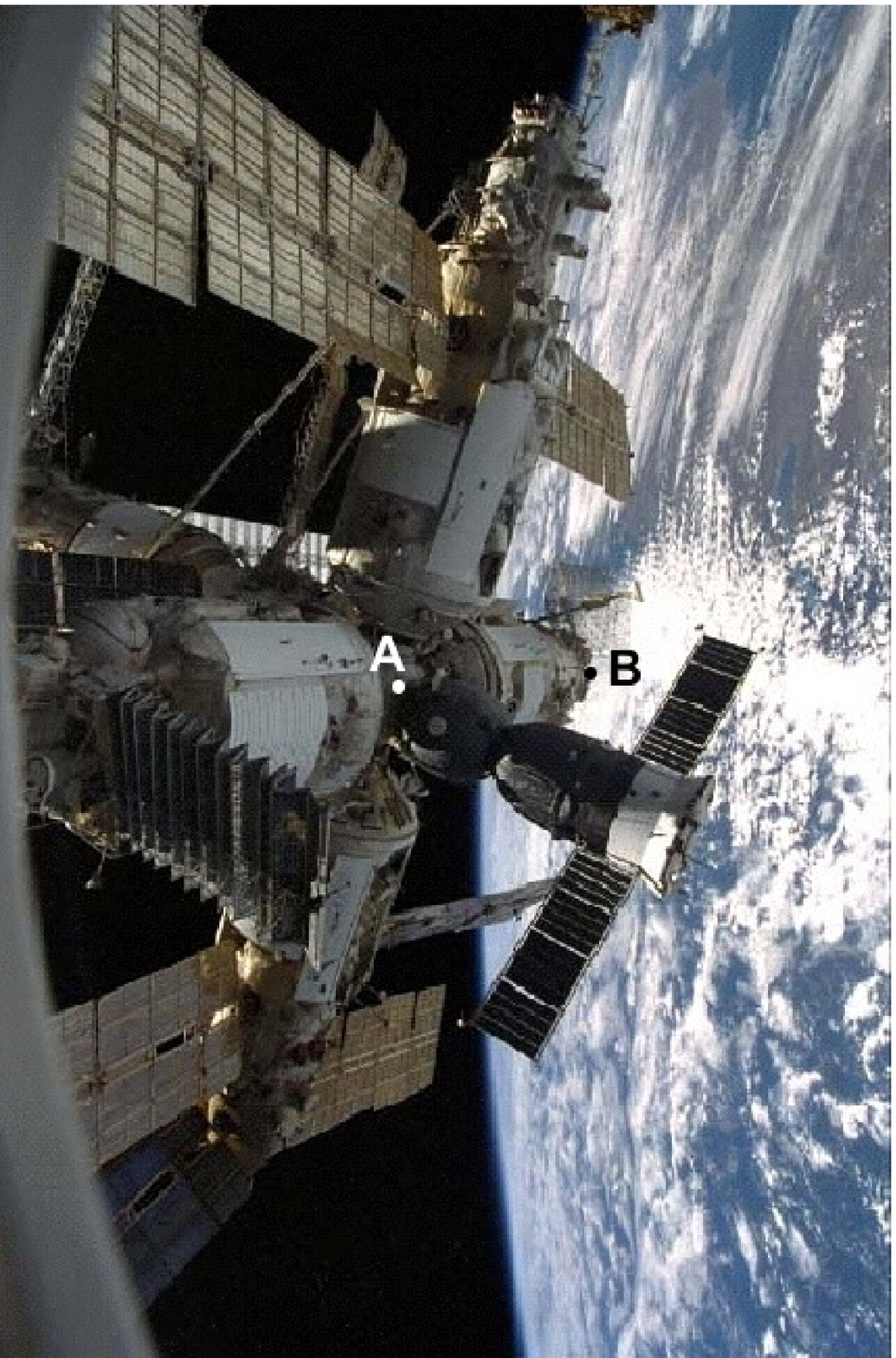}}
\hspace{.2in}
\subfigure{
\epsfysize=3in
\raisebox{1.24in}{\epsfbox{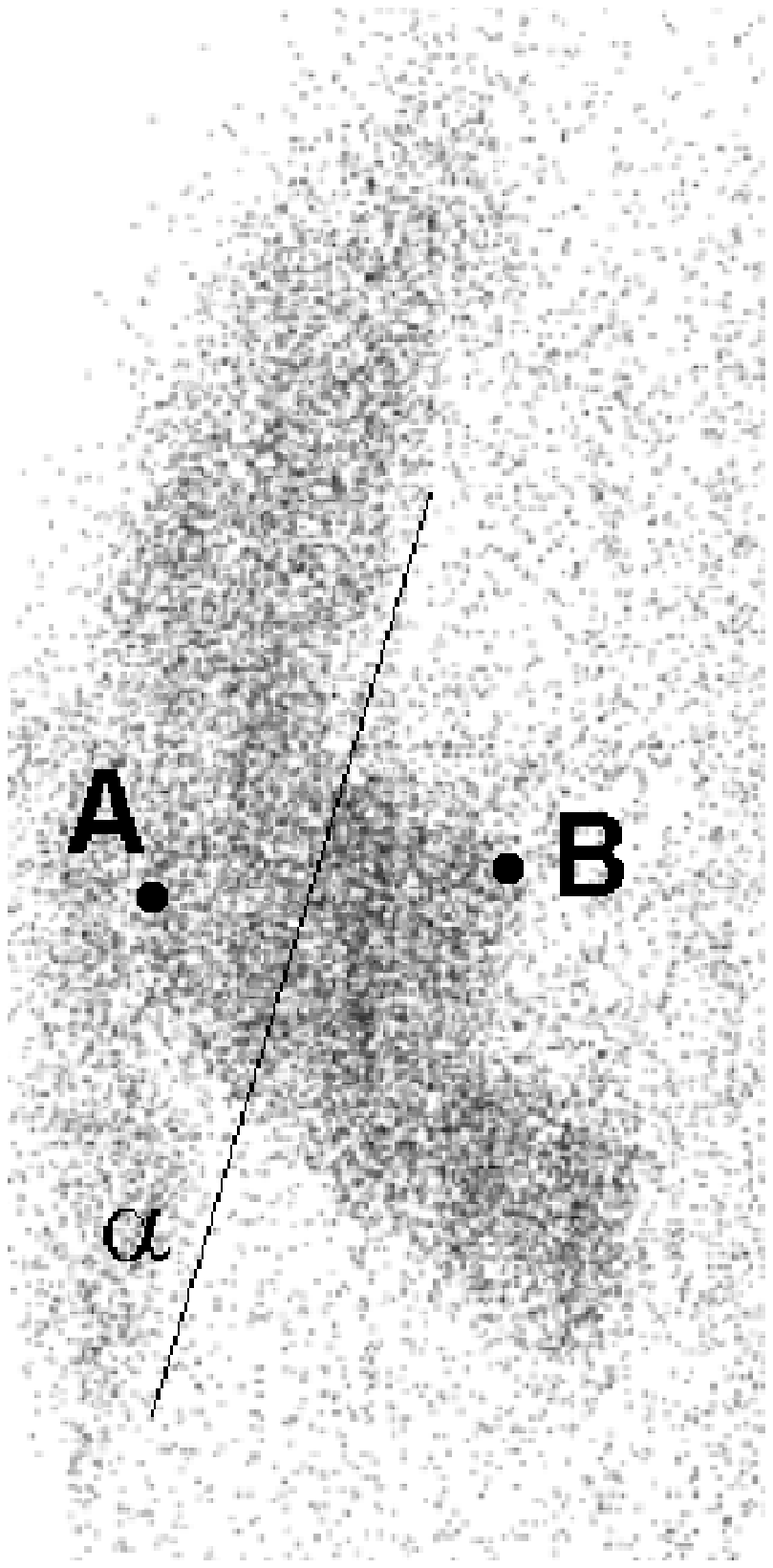}}}}
\caption{The picture on the left shows the space station 
\emph{Mir} as seen 
from a porthole in a SpaceHab module in the back of the payload bay 
of \emph{Endeavour} during \mbox{STS-89}. Visible are the \emph{Spektr}
(pointing upwards), \emph{Priroda} (pointing to earth), \emph{Soyuz} 
(black and white, pointing to right), \emph{Kvant-2} (pointing down), and
\emph{Kristall} (pointing to left) modules.
The picture was taken  from a 
location similar to that of \mbox{AMS-01} relative to \emph{Mir}. 
The image on the right is a zoom of the projection plot for $Z=-1$
events. Note the striking correspondence if the solar panels on the 
picture are ignored. The points $A$ and $B$ and the angle 
$\alpha$ are discussed in the text. Image on left is courtesy of 
\mbox{NASA} and taken from \cite{Mir_Shuttle_WWW}.}
\label{fig:mir_from_endeavour}
\end{figure} 

A one-to-one 
correspondence between the excess seen in fig.~\ref{fig:zn_nc_down}
and the physical layout of the \emph{Mir} space station was found.
Fig.~\ref{fig:mir_from_endeavour} shows
a picture of \emph{Mir} taken from a porthole of a SpaceHab 
module at the rear of the payload bay of the shuttle \emph{Endeavour}
when it docked with \emph{Mir} during an earlier mission
\mbox{(STS-89).} The SpaceHab porthole during this flight was in a 
similar location as \mbox{AMS-01} during the \emph{Discovery} flight 
docking phase; hence the view in the picture is
similar to the view of \emph{Mir} from \mbox{AMS-01}. Also shown is
a zoom of the excess events region in the projection plot.
The correspondence between the two images
is striking if the solar panels in the picture are ignored. We clearly
see the \emph{Soyuz} spacecraft projecting to the lower right, as well
as the \emph{Priroda}, \emph{Spektr}, and \emph{Kvant-2} modules. 
The solar panels presented much less material for spallation, and 
$\mu^-/\pi^-$ emission from them are not visible in the projection plot, as
will be shown later.

As a check whether we are indeed seeing \emph{Mir}, 
two points were chosen by inspection on the projection 
plot that appear to correspond to the node of the \emph{Mir} station,
labeled point $A$, and the top (furthest away) part of the 
\emph{Priroda} module, labeled point $B$. 
The polar angles of the points relative to the \mbox{AMS-01} 
$z$-axis, as well as the angle between the shuttle's $x$-axis and 
\emph{Mir}'s $x$-axis, 
$\alpha$, were estimated from the projection plots. These quantities 
were also computed independently using the dimensions of the \emph{Mir} 
modules and the space shuttle, taken from \cite{Mir_Shuttle_WWW}. 
The estimates of $A$, $B$, and $\alpha$ from the projection 
plots are compared with the computed values in 
table~\ref{tab:proj_other_comp}. They
are in excellent agreement, leading us to conclude that the excess 
seen in the projection plot is indeed part of the \emph{Mir} space 
station.

\begin{table}
\label{tab:proj_other_comp}
\centerline{
\begin{tabular}{|l|c|c|}
\hline
Quantity & Projection Plot & Specifications and Images \\
\hline
Polar Angle of $A$ & $54.5 \, \pm 2.0^{\circ}$ & 
$51.5 \, \pm 1.5^{\circ}$ \\
\hline
Polar Angle of $B$ & $67.6 \, \pm 1.6^{\circ}$ & 
$66.0 \, \pm 1.5^{\circ}$ \\
\hline
$\alpha$ & $25.7 \, \pm 1.6^{\circ}$ & $26.0 \, \pm 1.6^{\circ}$ \\
\hline
\end{tabular}}
\caption{Comparison of measurements from projection plots to 
engineering specifications from~\cite{Mir_Shuttle_WWW}. The errors in the 
second column are the
estimated uncertainty in locating points $A$ and $B$ by eye. The errors in
the third column 
are the uncertainties due to the location of \mbox{AMS-01} relative to
\emph{Mir}.}
\end{table}

The detected $\mu^-$ and $\pi^-$ flux is a complex convolution of several 
energy and direction-dependent functions: the time-averaged incident
cosmic ray flux, the material
distribution and composition of the \emph{Mir} space station, 
the $\pi^-$ production cross-section, the 
survival probability of the $\pi^-$ and $\mu^-$, the detector
acceptance, and finally the 
reconstruction software efficiency. The evaluation of this complex 
expression is beyond the scope of this paper; however, a simple estimate
of the incident cosmic ray energy can be made from the data as a 
consistency check.

The average reconstructed ``ambient'' event rate from the region covered by  
\emph{Mir} when the shuttle is not docked with it is measured to be 
$0.27\,\mathrm{s}^{-1}$ and $0.010\,\mathrm{s}^{-1}$ for \mbox{$Z=+1$} and 
\mbox{$Z=-1$} 
events respectively. The rate for $Z=-1$ events in the same region increases 
to $0.064\,\mathrm{s}^{-1}$ 
during the docking phase, hence the events from \emph{Mir} dominate in this
region during the docking phase and occur at a rate of 
$\approx0.06\,\mathrm{s}^{-1}$. This is 
$\approx20\%$ of the ambient \mbox{$Z=+1$} rate. 
If we approximate \emph{Mir} as a target of $10\,\mathrm{cm}$ 
thick aluminum \footnote{This is substantially thicker than the hull of 
\emph{Mir}, but accounts for the equipment inside the hull.} 
and the inelastic cross-section for protons on aluminum as 
$45 A^{0.7}\,\mathrm{mbarn}$, where $A=26$ for aluminum~\cite{mois97}, then
it can be shown that $\approx25\%$ of cosmic ray 
protons will interact inelastically
with \emph{Mir}. Although $^4\mathrm{He}$ nuclei 
compose only 20\% of the cosmic ray flux at these energies, $\approx50\%$
will interact with \emph{Mir}, at a rate $50\%\times25\%=13\%$ that of 
the proton flux.
Hence, the rate of cosmic ray $^4\mathrm{He}$ and proton interactions in the
region covered by \emph{Mir} is $25+13\approx40\%$ that of the ambient 
proton flux 
from the same region. This implies an observed, average $\pi^-$ multiplicity 
of $20/40=0.5$ per interaction, assuming the detection efficiency for 
$Z=-1$ and $Z=+1$ particles are the same, and that 
all produced $\pi^-$ reach \mbox{AMS-01} as $Z=-1$ particles.
This multiplicity requires an incident momentum for the cosmic ray protons
and $^4\mathrm{He}$ of $\approx5\,\mathrm{GeV}/c$ per nucleon~\cite{gazd95}. 
This is consistent with the  most probable momentum
for protons that varies between $1-15\,\mathrm{GeV}/c$ during the orbit due to 
the latitude-dependent geomagnetic cutoff~\cite{agui02}.
 
The solar panels can be approximated as $1\,\mathrm{mm}$ of silicon with 
$A=25$, hence their spallation $\pi^-$ production can be approximated as 
$1\,\mathrm{mm}/10\,\mathrm{cm} = 1\,\%$ that of the
rest of \emph{Mir}. Since the observed $Z=-1$ flux from \emph{Mir} is 6 times
that of the observed ambient cosmic ray electron flux, the contribution of
the solar panels has to be $6\,\%$ that of cosmic ray electron flux. 
Such a small contribution is difficult to detect and explains why the solar 
panels appear invisible on the projection plots. 

% ----------------------------------------------------------------------------
\section{Conclusion}
\label{se:conclusion}

The \mbox{AMS-01} experiment detected a $\mu^-$ and $\pi^-$ flux from
cosmic ray nuclei interacting with the \emph{Mir} space station. The 
precision tracker allowed the arrival directions of cosmic rays 
to be binned such that an image is generated on which 
individual \emph{Mir} modules can be distinguished by their flux of 
short-lived $\mu^-$ and $\pi^-$. 

During the \mbox{AMS-02} experiment we expect parts of the ISS and support
vehicles to be 
within the detector's field of view. This result shows that we can use
the data directly, without resorting to expensive and less reliable 
simulation, to identify these parts.
Using a simple graphical test based on this imaging technique, 
affected regions in the \mbox{AMS-02} field of view may be removed from 
sensitive analysis.

% ----------------------------------------------------------------------

\section{Acknowledgements}
\label{se:acknowledge}

The success of the first AMS mission is due to many individuals and  
organizations outside the collaboration. 
The support of NASA and the U.S. Dept. of Energy has 
been vital in the inception, development and operation of the experiment.  
The interest and support of Mr.~Daniel S.~Goldin, former NASA  Administrator,
is gratefully acknowledged.  The dedication of Dr.~John O'Fallon, Dr.~Peter 
Rosen and Dr.~P.K. Williams of U.S.~DOE, our Mission Management team, 
Dr.~Douglas P.~Blanchard, Mr.~Mark J.~Sistilli and Mr.~James R.~Bates, 
NASA, Mr. Kenneth Bollweg and Mr. T. Martin, 
Lockheed-Martin, the support of the space agencies from Germany (DLR), 
Italy (ASI), France (CNES) and China (CALT) and the 
support of CSIST, Taiwan, made it possible to complete the experiment on time. 

We are most grateful to the \mbox{STS-91} astronauts, particularly Dr. 
Franklin Chang-Diaz who provided help to AMS during the flight.

% ------------------------------------------------------------------------

\bibliographystyle{elsart-num}
\bibliography{main}

% -----------------------------------------------------------------------

\end{document}